# Tock: From Research to Securing 10 Million Computers

Leon Schuermann
lschuermann@princeton.edu
Princeton University

Brad Campbell
bradjc@virginia.edu
University of Virginia

Branden Ghena
branden@northwestern.edu
Northwestern University

Philip Levis
pal@cs.stanford.edu
Stanford University

Amit Levy
aalevy@princeton.edu
Princeton University

Pat Pannuto
ppannuto@ucsd.edu
University of California, San Diego

## Abstract

Tock began 10 years ago as a research operating system developed by academics to help other academics build urban sensing applications. By leveraging a new language (Rust) and new hardware protection mechanisms, Tock enabled "Multiprogramming a 64 kB Computer Safely and Efficiently". Today, it is an open-source project with a vibrant community of users and contributors. It is deployed on root-of-trust hardware in data-center servers and on millions of laptops; it is used to develop automotive and space products, wearable electronics, and hardware security tokens—all while remaining a platform for operating systems research. This paper focuses on the impact of Tock's technical design on its adoption, the challenges and unexpected benefits of using a type-safe language (Rust)—particularly in security-sensitive settings—and the experience of supporting a production open-source operating system from academia.

## 1 Introduction

This paper is about the design, evolution, and deployment of Tock, a secure embedded operating system (OS) written in Rust. Originally a research OS intended to explore how the Rust language might improve embedded system security, today, Tock runs on root-of-trust hardware in data-center servers, securely boots laptops, and powers automotive as well as space systems. At the same time, Tock has remained a platform for OS research.

This paper describes our experiences evolving Tock, from its early days as a research system into an open-source project deployed on tens of millions of computers (Figure 1). We hope these experiences can provide valuable insights to systems research. First, Tock's distinctive constraints, from being a security-focused OS to enabling multi-programming in highly resource-constrained systems, force it to make a set of interesting and unconventional design decisions and tradeoffs. Second, Tock is one of the first OS kernels written entirely in Rust, before Rust was embraced by OS researchers [27] and practitioners [11, 12, 33]. We encountered unexpected challenges and opportunities from using Rust, which are relevant to the broader systems community. Third, Tock reached practical adoption and widespread deployment, often in security-critical applications, shepherded mostly by academic contributors. We reflect on how it achieved this, and the relationship between academic research and an open-source project.

We revisit Tock's original design in Section 2. Many of the original goals and constraints that informed the design of Tock remain relevant today. For example, its use of Rust to provide isolation and least privilege with virtually no CPU or state overhead remains an integral benefit for Tock. Similarly, eschewing the single protection-domain design of most embedded OSes in favor of a hardware-mediated separation of kernel and applications has proven important for security, but also more generally for supporting legacy applications in C, as well as for managing product development that spans multiple teams. Much of the early revisions of the OS addressed the challenges of using Rust to enable this design, and many of our early decisions remain today.

Section 3 explores how Tock had to evolve since its first release. As both development of Tock progressed and Rust matured, we discovered that some of the design of Tock's internals were incompatible with Rust's soundness requirements; resolving this required substantially redesigning the system call ABI, main kernel loop, and Tock's kernel exten-sion abstraction ("capsules"). An expansion in Tock's application domain focus drove these discoveries and changes. The OS was originally designed to support low-power urban sensing systems and wearable electronics. Very quickly after its first release, however, Tock attracted interest from companies and non-profit organizations that developed hardware root-of-trust chips. These devices, often resource-constrained microcontrollers, are the foundation for all computing system security. They store keys, verify boot images, and are the first step in verifying that a system has not been compromised. For these uses, the benefits of a kernel written in Rust rather than C were obvious. Furthermore, new developers strongly pushed for better support of Rust userspace applications.

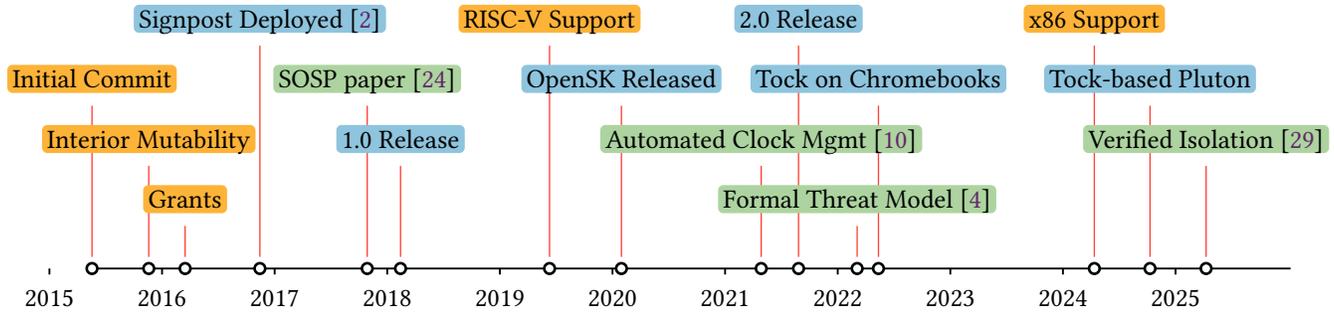

Figure 1: Timeline of Tock development, deployments, and research. Tock was originally developed to provide a secure multi-programming environment for urban-scale sensor network platforms. Since then, it has been adopted by users in other domains, primarily for hardware root-of-trust devices, which in turn motivated additional developments; these include a revised system call interface, a formal threat model, stronger isolation guarantees, and support for additional CPU architectures and hardware protection mechanisms.

In Section 4, we reflect on the various ways that Tock is able to use Rust's strong and expressive types to enable guarantees beyond memory isolation. Over the last decade, Tock has evolved type-based constructs for enforcing calling conventions that entirely prevent common driver bugs, for enabling memory sharing across multiple layers of complex software, and for generalizing hardware virtualization layers across platforms—all without dynamic memory allocation. These improvements can guide future OS kernel developers who are considering using Rust.

Over time, the amount of `unsafe` code in the core Tock kernel has remained steady, despite significant additional features (see Figure 5). While Tock always outlawed the use of `unsafe` language features in drivers, Section 4 further discusses how Tock refined its interfaces and design to contain and minimize `unsafe` usage. Today, `unsafe` is only used in a few limited places that must interact with hardware or software that is not subject to Rust types (e.g., hardware MMIO registers and the process boundary) and to create capabilities that limit access to sensitive kernel interfaces.

Despite the benefits of Rust, our experience has also highlighted the challenges of retaining language safety invariants when programming against both hardware and applications that are not beholden to language constraints (Section 5). The Tock kernel, like most embedded systems, is entirely event-driven with a single stack. Rust's memory model, however, is inexorably tied to threaded, synchronous execution: lifetimes correspond to stack frames, and borrows last over the duration of a function call. This makes upholding Rust's aliasing and mutability invariants when implementing an asynchronous system call boundary challenging, ultimately requiring a major redesign in Tock. Correctly interfacing with DMA hardware is similarly challenging. Mechanisms such as interior mutability enable common kernel architectures including circular dependencies between different layers, but make Tock less compatible with other embedded software developed in Rust. We share these insights to articulate how building an OS in Rust necessitates particular design considerations not present when using other languages.

Finally, we revisit questions around Tock's widespread adoption in Section 6. Most research systems are not adopted. Fewer still remain open source, and many that do have primarily been transitioned out of academia to a dedicated startup company or a single established organization. The Tock open-source project remains under the primary stewardship of academics and much (though not all) of the development, especially of core and security-sensitive functionality, is done by faculty and students at universities. Doing so has posed challenges: Academic pressures incentivize a focus on novel research, rather than engineering, while industry pressures incentivize meeting product timelines rather than making contributions available upstream. Urgent product requirements, such as fitting existing code into limited resources, may conflict with long-term thinking about sustainable and maintainable design decisions.

We argue that despite its challenges, this model has allowed the project to remain a realistic platform for research in operating systems, security hardware, sensor networks, and embedded verification, while facilitating collaboration across companies that would not have happened otherwise. As other research platforms gain traction, the model Tock has used may serve as an option for translating academic research into wider adoption.

Ultimately, while this long term collaborative effort has influenced and enhanced the evolution of Tock, the project has been able to remain focused on the original research questions and design.

## 2 Tock's Original Design

Tock's original design was targeted towards multi-programmed and battery-powered or energy harvesting sensor network platforms. The main constraints in these settings were robustness and security in the face of incorrect or nefarious kernel drivers and applications and, crucially, energy consumption. Tock also needed to run on and provide portability across a diversity of hardware platforms, each with different accelerators, sensors, wireless protocols, and physical interfaces.

Tock targeted a class of computers characterized by:
1. 32-bit Cortex-M architecture,
2. ~100kB of RAM and less than 1MB of executable nonvolatile storage,
3. no virtual memory, but primitive memory protection,
4. an abundance of on-chip hardware peripherals and accelerators, and
5. low-level access to a wireless radio.

For example, the first deployment of Tock was Signpost [2], a network of sensor network platforms spread across two University campuses. Each deployment unit consisted of a number of independent but interconnected boards, each with a microcontroller and a selection of different sensors (temperature, pressure, motion, sound, light, wireless spectrum, etc), as well as WiFi, LoRa, and Bluetooth radios to communicate with each other, pedestrians, and data collection servers. Each installation ran a number of common applications (e.g., server and peer-to-peer communication, inter-board coordination, etc) despite quite different hardware configurations, as well as a number of location- and hardware-specific applications (e.g., vehicle detection, air quality monitoring, etc). Finally, Signpost installations were solar-powered and, thus, power-constrained.

From these application and hardware-constraints, we derived many of Tock's initial core design decisions (Figure 2). For instance, the diversity of applications and sensors present in Signpost meant that Tock should have the ability to run multiple, isolated (potentially buggy or malicious) and concurrent applications, and its kernel must be highly extensible to support new microcontrollers and peripheral devices. The power constraints of Tock's target platforms, in turn, required a design that maximized the time that it can spend in hardware sleep states. We achieved this through a fully asynchronous kernel design, and extending this asynchronicity into userspace applications through its system call interface. Finally, the wide-spread nature of these deployments required them to be dependable, and avoid fate-sharing across applications and drivers, and between applications. To this end, we adopted a heapless kernel design with strict separation of application-related state.

These design choices mostly departed from typical embedded frameworks common in low-resource microcontrollers [3, 14, 36]. Most production embedded systems are designed for extensibility, but each component (e.g., a device driver, virtualization layer, or application) has unbridled access to the system and memory—there is no isolation. Typically all application functionality is compiled and delivered along-side the rest of the system and cannot be replaced dynamically. Because most systems do not separate applications from a kernel, there is no system call interface and drivers are free to expose arbitrary interfaces, which often mix and match asynchronous and blocking I/O. One commonality is avoiding heap allocation, which is typically discouraged in embedded systems due to the difficulty in reasoning about resource exhaustion, but is common in many other operating system kernels to support dynamic application demands.

A combination of a newly available type-safe language and increasingly common support for hardware isolation in low-resource microcontrollers enabled Tock's stark departure from related systems while still minimizing resource overhead and supporting power-efficient operation. For instance, it embraces Rust's strong type- and memory-safety properties to provide isolation between different *capsules* (Tock's kernel extensions), without using costly and scarce hardware memory protection resources. This design was influenced by prior research systems that used type-safe languages for kernel extensions, including Spin's use of Modula-3 [7] and Singularity's use of Sing# [20]. Rust's ownership types provided an opportunity to leverage such a language without the resource or performance overheads of dynamic memory allocation and garbage collection.

However, we discovered that Rust also places some unexpected restrictions on the design and architecture of our operating system. Specifically, its concepts of ownership and borrowing present challenges when composing an operating system out of many interconnected subsystems. This, in turn, influences Tock's programming model.

Over the remainder of this section, we want to focus on Tock's core design decisions. Starting with the challenges of interior mutability and how it influenced Tock's kernel architecture, we then illustrate its mechanisms for kernel extensibility and how it maintains application-specific state in its heapless design. We show how Tock uses both language-level isolation for kernel isolation, and enables conventional multiprogramming through processes with hardware isolation. And finally, we illustrate Tock's asynchronous kernel and system calls.

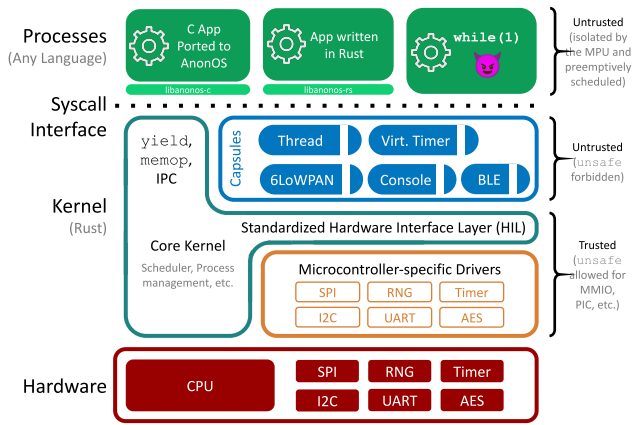

Figure 2: A Tock system is divided into four categories of components. Processes are hardware-isolated, preemptively scheduled, written in any language and are un-trusted. They interact with the kernel across a well defined and narrow system call interface the kernel. The kernel is composed of a common core, trusted hardware-specific adaptors, along with Capsules—co-operatively scheduled semi-trusted extensions isolated using the Rust type-system that are access process memory and hardware through narrow, restrictive, interfaces.

## 2.1 Interior Mutability

A defining feature of Rust is its ownership model. Any Rust object can either be *moved*, causing the original owner to lose ownership, or *borrowed* temporarily (by taking a reference). For any borrowed object, there can be at most one unique mutable reference, or one or more shared immutable references, and never both. This restriction is fundamental to retaining type-safety [17].

Many of Rust's safety advances build on these basic concepts. However, this model complicates building an operating system. To illustrate this, consider two operating system components, a file system and an underlying storage driver: When the file system performs a write, it modifies its own internal state, and calls into the storage driver through a reference. The storage driver then modifies the storage subsystem's internal state, performs the operation, and upon completion issues an asynchronous callback to the file system driver, again through a reference. Such a conventional design cannot naïvely be implemented in Rust. This is because *both* drivers modify their own internal state (and thus must invoke each other through a unique, mutable references), but they form a circular reference—the compiler cannot verify uniqueness of mutable references in this case.

This presents a design tradeoff. One possible approach is to use message-passing for interactions between drivers. In this model, all OS components would be owned by a single broker. Components can exchange data through this broker by addressing themselves through non-reference IDs. This avoids circular references. Yet, it replaces straightforward function calls between components with a central manager which introduces complexity and overheads.

The second possible approach is based on *interior mutability*. An object with interior mutability can change its state even when referenced through a shared, immutable reference. This allows both components to reference each other using shared, immutable references, which do not impose any restrictions on circular references. This permits conventional references between components. The cost for using interior mutability is twofold: certain state mutations add an additional runtime check, and additionally developers must be careful to ensure that reentrant calls into a single driver do not break consistency.

Ultimately, Tock settled on using interior mutability to provide a more explicit program structure with traditional-looking (circular) references shared between components.

## 2.2 Extending the Kernel With Capsules

*Capsules* represent Tock's primary kernel extension mechanism. We use capsules to implement system call drivers, *virtualizers* (components that multiplex multiple independent users on single hardware peripherals), and other non-privileged kernel infrastructure.

Making capsules special is our use of Rust's memory- and type-safety guarantees for isolation. We implement capsules as Rust structs in crates (Rust's compilation units) where use of `unsafe` Rust code is forbidden. This limits each capsule to accessing its own state and interacting with the rest of the system through explicit, safe APIs made accessible when instantiating the capsule.

This approach provides fine-grained isolation enforced at compile-time with virtually no runtime overhead. Capsules are trusted only to uphold *liveness* (e.g., to not busy-wait or `panic` the kernel). We implement the majority of unprivileged code in Tock through capsules; only low-level chip support code, MMIO peripheral drivers, and core kernel infrastructure need to be implemented as other, privileged, non-capsule code.

## 2.3 Hardware-Isolated Userspace Processes

Tock can also run preemptively scheduled and hardware-isolated userspace processes. Originally, these processes were intended to enable disentangling complex state machines typically found in embedded systems into distinct applications without requiring re-writing all code into Rust. Our vision was to exploit mostly idle microcontrollers for new functionality by enabling multiprogramming with multiple independent applications written in different programming languages (e.g., C and Lua) running on the same underutilized embedded platform. This eventually led to a formal threat model for applications, specifying that application data must remain secret, even from the kernel unless explicitly revealed by the application. We achieve this through the limited, safe APIs available to capsules. Additionally, applications cannot deny service to each other at runtime—in other words, they are *mutually distrustful.*

### 2.4 Dependability with a Heapless Kernel and Grants

Enabling multiple applications with high availability on a memory-limited hardware platform introduces a key challenge for dependability: Applications can exhaust memory by continuously requesting resources from the kernel. Specifically, kernel components often need to maintain state *related to*, but *inaccessible to* userspace applications. When using a shared kernel heap for such allocations, one application can exhaust these resources and infringe on the availability of other applications.

Tock prevents this using a key design feature, *grants*: the kernel does not feature a global heap, relying solely on static memory allocations and its stack. All dynamic memory allocations are instead placed in the *process'* memory region, and made inaccessible to userspace through the platform's memory protection mechanism. This ensures that processes can only exhaust their own memory, and that application-related state does not outlast the process lifetime.

### 2.5 Asynchronous All the Way Down

To meet its multiprogramming and low-power constraints, we adopted a fully asynchronous kernel architecture for Tock. This allows it to embrace asynchronous hardware (such as DMA-enabled peripherals), multiplex between work of different applications, and return to low-power hardware sleep states when there is no work left to do. Tock uses split-phase asynchrony, as in TinyOS [23], rather than message passing more common in microkernels that can rely more readily on dynamic memory allocation [1, 18–21, 25, 32, 35].

Notably, Tock was built before Rust gained native support for asynchronous programming models and instead uses a single-stack architecture with circular references and callbacks between kernel components to signal completions. Section 5.3 explores the implications of this.

However, Tock did not stop there: in addition to a fully asynchronous kernel, we also made the system call interface fully asynchronous. To perform any long running operation, such as printing a byte-buffer via UART, applications issue a sequence of calls to the kernel: they share a buffer of data with the kernel, register a callback (termed an "upcall") to be invoked by the kernel when the operation completes, and then request the operation to start. From this point onward, applications can perform any other task, or even start another asynchronous operation concurrently. Finally, applications can *yield* to the kernel to await completion, allowing the system to transition to low-power and the kernel to eventually invoke the registered callback.

## 3 Evolution

Today, the primary use of Tock (outside academia) is for root-of-trust hardware platforms including second-factor authentication (2FA) devices and security co-processors in laptops and servers. These settings share similar hardware constraints and application requirements as the original sensor network setting. However, they also differ in significant ways. These systems have both been a surprising match to parts of Tock's original design and required changes to others. In one case, Tock's generality proved too burdensome to square with product requirements and resulted in a hard fork. In another, it led a company, after using Tock for a few months, to design another custom Rust operating system for their server's hardware root-of-trust.

### 3.1 Hardware Root of Trust

Root-of-trust hardware devices are similar in many respects to sensor networks hardware. They typically have very little memory (tens or hundreds of KB) and use a single CPU core, typically 32-bit. They do not have virtual memory but do support single-address space memory protection. They include many specialized and sometimes proprietary accelerators and peripherals, especially for cryptography.

Their software requirements are also similar to the urban sensing networks Tock was originally designed for. Hardware roots-of-trust often run multiple, independently developed applications that need to be portable across hardware versions. They often need to support multiple application languages: while Rust is attractive for new services and applications, users also want to continue using their existing, highly tested and dependable C applications. Applications and the OS greatly benefit from being resilient to bugs in both applications and drivers.

Early adopters included two products from Google: OpenSK, a FIDO-compliant second-factor authentication device [15], and Ti50, a hardware root of trust for Chromebook laptops [16]. Later came vendors for root-of-trust hardware on datacenter servers, and other end-user device manufacturers.

Over several years of engagement and development, we adapted to multiple important differences between hardware roots-of-trust and urban sensing networks. Some of these were easily compatible with Tock's design, while others raised tensions. In this section, we focus on the most important differences, and how they changed Tock's design.

### 3.2 Complexity of Asynchronous Code

Tock was originally designed to support ultra-low power operation, a common requirement for wireless sensor networks and for the Signpost system. Prior operating systems and applications had established that asynchronous I/O provided substantial energy saving benefits because a single thread could execute many operations in parallel and so spend less time awake [22]. Therefore, Tock embraced both an event-driven, single-threaded kernel as well as an asynchronous system call interface.

Hardware roots-of-trust, however, are either not energy-constrained or, when they are, much less so than wireless sensor networks. A hardware root-of-trust on a laptop or server's energy use is insignificant in comparison to the larger system. USB-based 2FA devices have steady power through the USB port, and even wireless Bluetooth ones do not need to tightly optimize their duty cycle, as it is governed more by human reactions (seconds) than I/O timing (microseconds to milliseconds).

As a result, while the asynchronous kernel has allowed Tock to support many simultaneous applications and services with a single stack, in userspace it has been more of a hindrance than benefit for many users. Root-of-trust applications are typically sequential state machines. A simple synchronous operation such as "wait for a response with a timeout" can become a half dozen system calls, to allow a buffer, register two callbacks, issue commands, then wait for a callback.

This mismatch led one early adopter of Tock, Oxide, to discard it after a few months. They wanted to use Tock for the firmware on their baseboard management controller (BMC). However, they realized they had a very fixed set of applications, all of which relied on straightforward sequential state machines. They saw the complexity in the kernel to handle the interleaving of asynchrony, plus the complexity in userspace to convert this into a synchronous API, and decided to write a new synchronous OS in Rust, named Hubris [9].

A second early adopter of Tock, Ti50, forked early on and stayed involved for many years, but over time their participation has diminished as their fork and mainline Tock have diverged. They also had concerns with asynchronous system calls, but for a different reason: code space. Their MCU is RISC-V based, and relatively immature LLVM code generation for RISC-V can result in very expensive system calls. At the same time, they were on an extremely tight timeframe to ship their firmware. They forked Tock to introduce a blocking `command` system call. This allowed them to change a 4 system call sequence (subscribe, command, yield, unsubscribe) into a single call. While there was interest from other developers to introduce such a new system call into the mainline kernel, once Ti50 had forked they were not strongly motivated to drive it, and it fell by the wayside. Since then, we have introduced a new variant of the `yield` system call, described below, that simplifies creating synchronous libraries using the asynchronous system call interface.

Today, while some users of Tock are ambivalent about synchronous or asynchronous system calls, others see them as essential for their root-of-trust and automotive applications. Asynchrony, however, had an additional complication, which necessitated a complete redesign of the system call API. We learned that the asynchronous semantics of the original system calls were incompatible with a sound Rust userspace without a lot of unsafe code.

### 3.3 Rust in Userspace and Tock v2.0

Tock's original design was intended to allow userspace processes to be written in any language (including assembly), but almost all applications were written in C. Early non-academic adopters argued that they would like Rust in userspace as well. While Tock had a Rust userspace library, we soon learned it was unsound. Properly supporting a Rust userspace with minimal unsafe code required a redesign of the system call ABI, core kernel loop, and the capsule interface.

#### 3.3.1 Respecting Ownership

The challenge that emerged is that Rust's memory ownership semantics had to be expressed in the system call ABIs and enforced by the kernel itself, rather than entrusted to capsules. Both the `allow` (sharing a buffer) and `subscribe` (registering a callback) system calls allow userspace to share references to resources in their memory: data in the case of `allow`, and code in case of `subscribe`. The original capsule API simply validated supplied values, and then passed ownership over types wrapping these references to capsule code, Tock's semi-trusted kernel extensions.

The problem arose when userspace tried to revoke one of these buffers or function pointers, by calling `allow` or `subscribe` again. While the system call semantics said that a capsule was required to replace the old value with the new one, this couldn't be enforced by the compiler: a capsule owned these wrappers, and could always stash the old one away in a variable. A Rust userspace needs to be able to carefully track who holds on to references to which resources, and how long values live, but capsules presented an obfuscated interface that could hold on to references for arbitrary lengths of time. A semi-trusted kernel extension holding onto and using a userspace reference arbitrarily would compromise all of Rust's safety guarantees, which greatly reduces the value of a Rust userspace to security-critical applications.

#### 3.3.2 Swapping Semantics

Today, the `allow` and `subscribe` system call APIs have "swapping" semantics, similar to how TinyOS manages buffers [23]. When userspace invokes one of these calls, the call returns the previously shared buffer or function pointer. On the first call, the kernel returns a dummy empty holder value: a zero-length buffer for `allow` and a special "null" callback for `subscribe`. Userspace is responsible for ensur-ing that references remain live while the kernel holds them. Within the kernel, capsules are no longer responsible for managing these buffers or callbacks; they are held and managed by the kernel. Capsules can access them only through temporary Rust references in closures, which ensure they cannot take ownership over these types. This final change required completely rewriting the core kernel loop.

### 3.3.3 Tock v2.0

This redesign of the system call API (and ABI) meant existing applications would no longer work. We debated having both versions exist concurrently, but the code size cost was too large. This led to a clean break in Tock v2.0. In addition to the swapping semantics of `allow` and `subscribe`, Tock added a new variant of `allow` that granted the kernel read-only access to a userspace buffer. This turned out to be a must-have for root-of-trust applications: they would often store public keys in read-only flash memory, and want to pass these values into the kernel for cryptographic operations. Without `allow-readonly`, userspace had to copy these into RAM; otherwise, it would be possible for the kernel to inadvertently try to write to read-only memory and trigger an unrecoverable fault.

## 3.4 Process Loading

Tock's process architecture supports independent applications that the kernel can individually load, start, and stop. This was designed to support flexibility for deployments in the field (e.g., Signpost). In an early root-of-trust application, Tock was not the lowest layer of trusted software. Instead, a trusted bootloader verifies a signed image that includes Tock and a set of applications. While this required repackaging them together along with a signature, these changes were external to Tock and so required no changes to the kernel itself.

One of the root-of-trust use cases introduced a new security model, in which applications could be individually signed and replaced. This meant that the Tock kernel had to be able to run a series of cryptographic checks on process binary images before executing them. Cryptography implemented in hardware peripherals is asynchronous, so this forced the process loading sequence from a simple synchronous pass over the header and integrity checks into a multi-step state machine. Processes are first checked for structural/header integrity, then cryptographic integrity and authenticity, and finally runnability. This in turn required rewriting the Tock boot sequence from a piece of sequential code into an asynchronous state machine.

The transition to support signed applications had one major benefit and one large drawback. The benefit was that it paved the way to dynamically load and run new applications without rebooting, a feature some users had always put as desirable but not critical. As process loading was now an asynchronous state machine, this became much easier: all the system had to do was trigger the kernel to check the new process. The drawback was code complexity. For codespace-limited systems that didn't need dynamic process loading (e.g., the single signed image), the new approach added a lot of code which wasn't needed. The kernel therefore supports either approach, which can be configured at compile-time: the boot sequence for a build can either invoke the synchronous or asynchronous loader.

## 3.5 External Dependencies

One final way in which Tock has evolved is an issue that many secure Rust systems have struggled with: external dependencies. Because `unsafe` Rust code can perform arbitrary operations, including an external dependency requires that one trust it does not violate the safety guarantees the system depends on. Rudra, for example, demonstrated many places in which external dependencies could be used to break memory safety [6]. Initially, Tock took a very hard stance on this: the only external dependency allowed is the Rust `core` crate, part of its standard library.

In practice, though, third-party libraries provide significant useful functionality, much of which would be difficult or undesirable to re-implement or vendor. For example, root-of-trust use cases rely heavily on cryptography, often in privileged kernel drivers. In some cases, cryptographic primitives are provided by the hardware, but often higher-level protocols, such as stream ciphers, signature verification, or certificate validation must be done in software. Correct implementation of cryptography is security sensitive beyond just memory safety, and it is undesirable to re-implement ciphers or protocols or maintain implementations, especially when there are well-maintained or even formally verified third-party libraries [13]. Tock has made a specific exception to the external dependency rule for certain cryptography libraries that have a limited dependency tree and are robustly maintained.

However, using third-party libraries would be useful in more cases, such network and wireless protocols. Trusting third-party libraries remains an open research problem [26, 34].

## 4 Discovered Rust Opportunities

Rust's expressive type system has benefited Tock development beyond mere memory- and type-safety guarantees. We have been able to encode higher-level properties of both operating system and embedded system design using the type system, enabling compile-time checks against both invalid and unacceptable system compositions, improved buffer management, and expressive hardware interfaces.

## 4.1 Enforcing Correct Composition

Tock supports a range of hardware platforms and is designed to be configurable for a rich set of devices. We achieve this flexibility by dividing up system components into composable layers. For instance, a hardware timer peripheral implements a common interface that allows clients to interact with this peripheral in a hardware-agnostic way. Then, another layer can use this interface to *virtualize* the single timer resource for multiple clients.

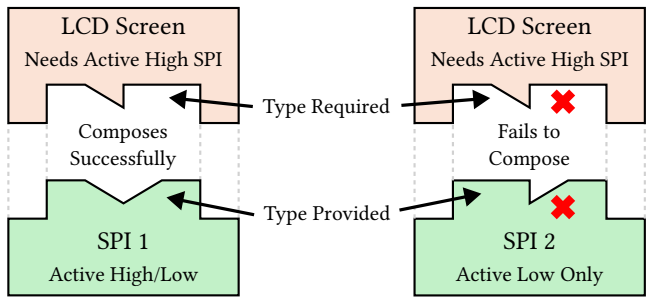

Figure 3: Enforcing layering requirements at configuration time. Higher-level hardware-agnostic drivers specify their needs and build on top of lower-level peripheral drivers which express their capabilities. Including these requirements in the Rust type system allows Tock to ensure configurations are valid at compile time.

However, this flexibility also comes with challenges: configuring a Tock kernel instance for a hardware platform requires correctly matching microcontroller-agnostic drivers, like a SPI-connected sensor, with the appropriate hardware-specific peripheral implementation, like a SPI controller. With this composition, even subtle mismatches in the assumptions and limitations of layers can result in (subtly) incorrect behavior. Drivers are commonly located in separate files, written by different developers at different times. Incorrectly assuming a certain driver stackup is valid is an easy, but challenging to debug, mistake.

Instead, we have found that we can use Rust's type system to validate correct composition at compile time. We do so by allowing drivers to express high-level aspects of the interface they provide through type annotations. Clients of these interfaces, in turn, annotate type-constraints to only accept drivers that provide compatible semantics.

For example, the SPI communication bus allows multiple external devices to communicate with a microcontroller using a shared bus. A separate "chip select" (CS) pin for each device indicates that software has selected that device to use the bus. Different devices expect the CS pin to be active-high, active-low, or configurable. This is microcontroller-agnostic information that a higher-layer driver can encode. Similarly, some SPI hardware supports only active-high pins, only active-low pins, or can configure the active CS level. Even within a single microcontroller, multiple SPI peripherals may have different such restrictions. This is information that the microcontroller-specific hardware driver should know. Correct composition requires the SPI driver to both support and be configured for the same active polarity as the peripherals it connects to (Figure 3).

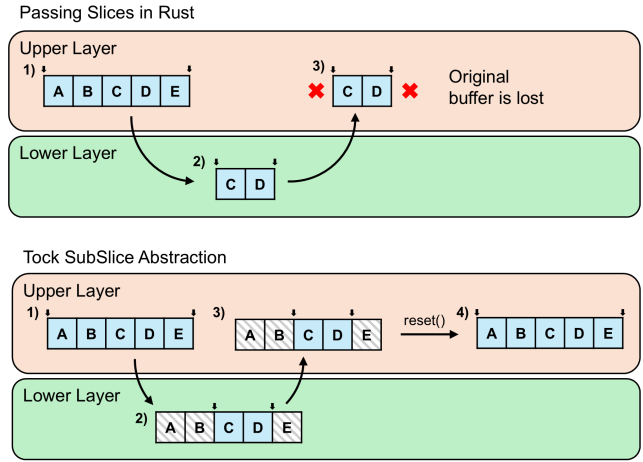

Figure 4: Tock's `SubSlice` abstraction allows kernel components to pass buffers back and forth, resize them at will, while retaining the ability to restore access to the complete underlying buffer.

Using template constants in Rust types we can express the capabilities of hardware drivers and the requirements of chip-specific drivers, and ensure they are configured correctly. Mismatches are caught at compile type through a type error.

### 4.2 Buffer Management

To support concurrency within the kernel with a single stack, internal Tock kernel APIs rely heavily on split-phase asynchrony. A split-phase operation has two parts: a method call to start the operation and a callback to signal completion.

Rust's buffer management complicates split-phase APIs. The canonical way to represent a buffer in Rust is with a slice, which is a pointer and length. Rust APIs typically pass slices as a temporary reference, whose lifetime expires when the function returns. Asynchronous operations, however, must be able to hold on to the reference until the operation completes. For instance, a DMA-capable radio needs to hold unique "ownership" over a buffer beyond the end of a `start_receive()` function call, until an asynchronous interrupt indicates that a full packet has been received.

Tock APIs allow subsystems to hold on to buffers longer than the duration of a function call by passing slice references with a `'static` lifetime. The `'static` lifetime is a special lifetime in Rust that means the reference is valid for the entire program duration (Tock allocates its buffers statically at boot). Because Rust's ownership rules prohibit multiple unique, mutable references, passing a `'static` reference to a driver makes it inaccessible to the caller of that driver; calling the method conceptually *moves* the buffer's ownership into that driver. The driver provides this eternal reference back to the client in the completion callback.

This approach of passing ownership of `'static` buffers across software layers, however, breaks down when trying to pass dynamic subsets of data. A caller cannot simply pass a subset of a buffer (i.e., increment the start pointer), because doing so would mean giving up ownership of this part of the buffer forever, and two subslices originating from the same buffer cannot later be re-merged into the original, larger one. Each layer must handle the complete buffer, so it can be returned in its entirety to the original caller. Early versions of Tock kernel interfaces tried to solve this problem by accepting both a slice as well as a separate offset and length describing the specific subset of the buffer to operate on (e.g., the fraction of the buffer to encrypt with a key).

Unfortunately, requiring interfaces to pass a slice, subset offset, and subset length proved cumbersome and error-prone. Tock introduced two mechanisms to help alleviate this problem. This first is a `SubSlice` type (Figure 4) that allows individual layers to access a buffer as if it was exactly the length of the valid data, without losing ownership of the entire buffer. The second mechanism uses `const` parameters to enable software layers to express buffer size requirements. These parameters are accessible at compile-time and allow higher layers to construct static buffers that are sufficiently sized for each of their dependencies.

### 4.3 Memory-Mapped I/O Abstractions

Microcontrollers conventionally use Memory-Mapped I/O (MMIO) to interface with hardware peripherals. Peripherals claim memory addresses where they respond to memory read or write events, typically triggering underlying hardware operations. Translating hardware specifications from a datasheet into raw memory addresses and packed structs that correctly map to the hardware's memory layout, and modeling access permissions for each memory mapped register is tedious and error prone.

Tock leverages the Rust macro system to provide a domain specific language for capturing MMIO specification. This high-level specification maps closely to typical datasheets, making it easy to translate them into code and avoid mistakes. The resulting data structure wraps each MMIO address in a unique type, which only exposes read or write operations supported at that address. These address-specific types can further define subtypes, `Field`s, that encode information such as valid enumerations as well as offset and length within a memory word. Now, bit-shifting code is automatically generated, rather than left to error-prone manual code.

```rust
// To call this function, the caller must
// have a ProcessMgmtCap instance to pass in,
// which is an empty marker trait that is
// otherwise unused and elided at runtime.
fn destroy_process(&self, _cap: &ProcessMgmtCap) {
    ...
}

// Capabilities can only be created during
// platform initialization. They can then be
// passed into privileged drivers.
struct MyCap;
unsafe impl ProcessMgmtCap for MyCap;
let privileged_driver = ProcessManagementDriver {
    cap: MyCap,
    ...
}
```

Listing 1: Tock uses marker traits to restrict access to sensitive APIs within the kernel, allowing untrusted capsules to gain special privileges, but *only* if trusted platform initialization mints and passes a compile-time-only capability. Attempting to call the API function without the capability will fail at compile-time.

### 4.4 Privileged APIs with Capabilities

Rust distinguishes between safe and `unsafe` code, and the compiler can reject `unsafe` on a per-*crate* (a Rust compilation unit) basis. This enables Rust code to have restricted APIs—marked `unsafe`—that can only be called in certain contexts. For an OS, this is particularly useful, as certain privileged functionality must be exposed (e.g., stopping a process), yet many kernel modules that interact with processes (e.g., implementing userspace timers) should never be able to invoke this functionality. However, while marking functions `unsafe` provides an enforcement mechanism, it provides only all-or-nothing access without context as to when access should be provided. Tock remedies this with a "capabilities" mechanism (Listing 1), built on the compiler-enforced `unsafe`, to create zero-sized types (hence, with zero overhead at runtime) to express restricted APIs and use the type system to mediate the permissions required to use those APIs.

## 5 Discovered Rust Pitfalls

While Rust has benefits for implementing an OS, it also introduces unique challenges for writing low-level code. Over the course of Tock's development, we found instances where traditional wisdom and practices for building OSes no longer applied, and could even lead to violating some of Rust's invariants required for safety and soundness.

### 5.1 Rust's Memory Model for an OS

Many of Rust's innovations around safety focus on restrictive assumptions around how memory is used. For instance, recall that a Rust value may have either multiple *immutable* shared references, or a single *mutable* reference, and never both. The compiler checks these invariants through its lifetime and borrowing mechanisms.

However, these rules are difficult to enforce in an operating system context. The notion of running processes, with application code not implemented in Rust, results in the compiler lacking visibility to enforce memory invariants. Alternative approaches like Theseus [8] avoid these issues by requiring all operating system components and applications to be written in Rust and benefiting from compiler checks. However, our experience shows this is not practical: real-world devices must be able to run legacy software written in C, and crucial benefits stem from process isolation.

Instead, Tock employs hardware memory protection mechanisms and multiple privilege modes to isolate untrusted userspace applications. Hence, Tock must bridge Rust's static, compile-time safety invariants onto this userspace–kernel boundary, either by weakening the set of invariants that Rust expects to be maintained, or by employing dynamic checks across interactions between kernel and userspace components. In practice, Tock uses both approaches to efficiently preserve Rust's soundness across interactions with untrusted code. We illustrate this by revisiting Tock's `allow` mechanism, used to exchange data between the kernel and userspace applications [24].

### 5.1.1 Mutably Aliased Application Memory

When an Tock application issues an `allow` system call, it passes a pointer and length to share a buffer with a kernel capsule. Although capsules run in kernel mode with access to all userspace memory, their restriction to using only *safe* Rust prevents them from accessing arbitrary memory. By explicitly *allow*-ing buffers, applications give these kernel components permission to read and write a specific section of their memory. In the kernel, we represent such a buffer through an `Allow` handle which, after checking that the issuing process is still alive, can be used to safely obtain a Rust *slice reference* to its underlying memory.

This mechanism is similar to how memory is shared between the kernel and userspace applications in other OSes. However, Tock's use of Rust presents some additional challenges that made our initial implementation unsound. First, applications can share the same buffer with two *allow* system calls, giving a kernel capsule two mutable references to the same memory. This breaks Rust's *aliasing XOR mutability* invariant. Writing to the first buffer may change the contents of the second, whereas the compiler assumes that the contents of the second remain stable. This is unsound, and such violations of Rust's aliasing assumptions have led to exploitable security bugs in practice [28].

To address this, ideally the Rust compiler would statically determine that two shared buffers can never overlap, but since applications are compiled separately this is not possible. Alternatively, Tock could reject overlapping shared buffers with a runtime check. However, this introduces unreasonable runtime overheads for the systems that Tock targets. Instead, we choose to reduce the invariants that Rust requires us to maintain for shared buffers. By changing their type signature from a mutable byte-slice (`&mut [u8]`) to a shared slice over interior-mutable *cells* (`&[Cell<u8>]`), Rust no longer assumes that the buffers contents remain stable for the duration that their references exist.

### 5.1.2 Violating Rust's Type Invariants

Sharing process memory with the kernel led to another issue around upholding Rust's strict requirements for soundness. Applications reclaim a shared buffer from the kernel by issuing another allow system call with a zero-length buffer. The kernel held a reference to this zero-length buffer, even if its address was null. However, this was unsound: Rust requires that the address of references must never be null. Based on this requirement, it performs an optimization known as *niche filling*: it can use uninhabited values of type to encode other information. For instance, given that references must never be null, a value of type `Option<&[u8]>` occupies the same space as a value of type `&[u8]`, as the `Option`'s `None` variant can be simply be encoded as a null-value. However, this means that Tock's practice of taking an arbitrary, user-supplied pointer as the address for a zero-length slice reference breaks Rust type-safety. Instead, we must use a dynamic, runtime check for this operation and use an artificial, non-null pointer for creating zero-length slice references shared with the kernel.

### 5.1.3 Emphasis on the `unsafe` Boundary

Bugs like these show that, while Rust can give developers confidence in the *safe* part of their programs, it also requires significant care to ensure that the *boundary* of safe code maintains the extensive set of safety-invariants that Rust imposes. In many cases, prior languages like C imposed fewer such invariants (e.g., concerning aliasing). And while the consequences of violations are often subtle and dependant on compiler optimizations, they can nonetheless introduce critical safety and security vulnerabilities in arbitrary, safe and unsafe parts of the codebase.

## 5.2 Lack of `unsafe` Guidance

As the previous section illustrates, explicitly separating safe and unsafe code is both a major feature, and requirement of correctly using Rust. However, during early stages of Rust's development there was little guidance on how and when to use `unsafe` in a large software project to ensure maintainability and correctness over time. Our experience shows that modularizing the OS and prohibiting `unsafe` in certain crates provides clear guidance for developers and encourages external contributions that are consistent with Tock's design. However, within crates that do permit `unsafe`, we have found that developers often use `unsafe` in new ways, particularly to support bespoke hardware features, and then layer implementations on top of the `unsafe` use. Over time, this has lead to numerous `unsafe` uses that are both hard to reason about and hard to remove.

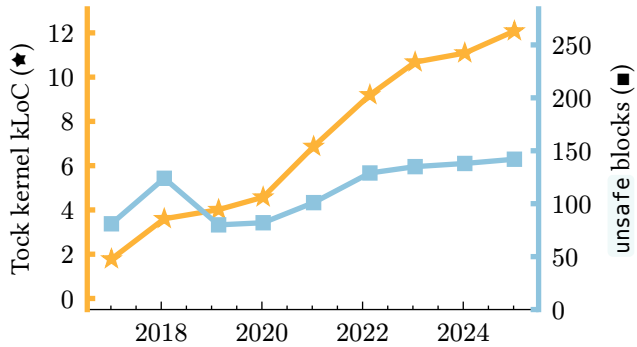

Figure 5: Increase over time in the size of the Tock kernel due to additional features and abstractions compared with the amount of kernel code that is `unsafe` Rust. While the kernel has grown significantly over a decade, `unsafe` blocks have remained steady.

Meanwhile, Rust has adopted a more narrow definition of *safety*, specifically focused around the notion that safe-code cannot exhibit undefined behavior [30]. This is distinct from other forms of functional correctness, for which Tock has at times used `unsafe`.

In hindsight, we should have created more abstractions around `unsafe` to both guide and support contributors (such as Tock's *capabilities*), and to limit the diversity of `unsafe` use in the kernel. While the project has managed to preserve a steady and low use of `unsafe` in the kernel, limiting its prevalence requires continuous maintenance work.

### 5.3 Using External Software Libraries

Despite a robust package management and dependency ecosystem, Rust has provided minimal benefit for Tock in terms of re-using existing drivers and software libraries.

A major challenge is the Tock kernel's asynchronous, single-stack design being incompatible with available libraries: most existing code is architected to be either blocking, incompatible with Tock's multi-programming model, or use Rust's `Future`s for asynchronous computations. However, we found that `Future`s have significant overheads compared to Tock's callback-based design.

Additionally, as described in Section 3, we are cautious to integrate external code into Tock as both the code and dependency tree is difficult to audit. Whether there is a safe way to integrate external code into an OS kernel remains an open question. We anticipate that additional tooling is required to analyze and continuously monitor external packages across package updates over time.

### 5.4 Verification Still Useful

Rust's type system provides a powerful foundation for, and incentivizes writing robust code. However, logic correctness remains entirely the programmer's responsibility. For complex subsystems with subtle correctness requirements we found that Rust does not substantially help avoid bugs. Two subsystems in particular–timer virtualization and memory protection–have presented numerous subtle logic bugs. We are exploring verification techniques integrated into Rust to improve correctness of these challenging subsystems.

## 6 Supporting an Open Source Ecosystem from Academia

Tock today is an open-source community led from academia. It never got swallowed up by tech-transfer to a single industrial user. It did not become a startup focused on providing support or development of the OS. Instead, it has walked the line of academic management and industrial deployment. Over three-quarters of the current leadership team is comprised of academics (with the remaining quarter coming from industry). Our experience maintaining an open-source, industry-deployed operating system from academia has presented several challenges and key benefits that we explore here.

### 6.1 Challenges of Academic Open-Source

Somewhat stereotypically, some of the largest challenges for Tock are "solved problems" with known implementations that still require substantial engineering effort to apply. Several times we saw academic focus fade once problems move past the hard and interesting stage. Engineers directed towards certain tasks can easily overcome this problem. This proves more difficult with independent contributors with primarily research motivations. Eventually, this can lead to gaps in support throughout the system.

As an example of this problem, one long-term need for the project has been a hardware-in-the-loop testing system. Supporting a diversity of architectures, chips, and specific board implementations requires frequent testing on real hardware platforms. In practice, we would find that some lesser-used platforms would be non-functional for weeks or months before someone would stumble upon this issue. While an automated hardware testing framework was a clear solution, the process of creating it was an enormous engineering effort without clear research questions. Meanwhile industry adopters were interested in using the framework, but were not incentivized to create a general-purpose system for testing hardware they did not use. After years of need, a framework was eventually developed by contracting the effort to a full-time engineer.

Collaboration between academia and industry can also face typical timeline mismatches. Faced with product deadlines, adopters have feature needs that cannot wait for academic consideration and discussion. For an open-source project, this tends to result in forks or custom downstream additions to the OS. To some degree, these are expected and acceptable. The design of Tock intentionally separates components so that they can be replaced if necessary. But downstream changes implemented by industry adopters do not necessarily pay back to the broader community. On several occasions we have found ourselves recreating features that had previously been implemented by industry users but had never been cleaned up and shared with the community.

## 6.2 Benefits of Academic Open-Source

Being rooted in academia has also provided important benefits for Tock. Sheltering in academia allowed Tock to survive its first few years of development. The time before initial adoption by external users is a challenging period for open-source projects. They miss out on the energy and direction provided by adopters. For OSes in particular, a challenge is that an early OS is only capable of supporting a few platforms with basic functionality. Through addition of drivers and platform support, it takes time to build up to a critical mass. Having to prove its value for startup funding during this initial period would have been very challenging for Tock. Instead, Tock operated for years with essentially zero funding. This gave us room to experiment and build while being "funded" by our day jobs in academia.

Later in the life of Tock, after we had established users and support, the academic core of Tock gave us a long-horizon view separated from production deadlines. What is internally useful and expedient to implement for one user is often not good for other users or potential future adopters. In the context of developing systems in Rust, tight timelines can increase the desire to get around the compiler via the `unsafe` escape hatch. This trades something that seems to work fine for now for future soundness concerns. Instead, the core development team on Tock has felt empowered to take the time to get "get things right".

The updates to the system call API, mentioned in Section 3.3, are a good example of this. The need for a revised syscall interface to support a Rust userland was first made apparent in early 2019. Other issues with the system call interfaces were collected and an official start to the 2.0 version of the kernel was begun a year later in early 2020. The first draft of the documentation for the revised syscall interface was created in mid-2020. The release of Tock 2.0 finally occurred in mid-2021. The revised design was given plenty of time to be discussed and considered before final deployment, and we were in no rush to complete it as major breaking changes in an OS kernel are a rare occurrence. Being insulated from production deadlines and startup funding cycles provided time and patience required for deep, technical discussions and considerations.

An added benefit of being hosted in academia is bringing undergraduate students into the project. The students gain training in real-world systems engineering, while Tock gains engineering effort, and there are a nearly-endless supply of drivers which can be added to the OS but are not immediately critical for any particular users. Many features were developed by undergraduate students interested in gaining experience.

For graduate students Tock has proved to be an excellent source of hard systems considerations that are still fairly constrained. Over the years, grad students have used Tock for experimentation with automatic clock configuration to save energy [10], for isolation of existing C libraries within a Rust system [31], for using the Rust type system to automatically enforce buffer size requirements at compile-time, and to reduce code size for Rust systems [5]. A recent collaboration includes grad students working on code verification systems who were interested in the constrained interfaces and security guarantees provided by Tock [29].

## 7 Conclusion

After a decade of development, multiple major releases, thousands of commits and pull requests, several developer meetings, multiple graduated Ph.D.s and new faculty members, and now tens of millions of active devices, this paper shares the evolution of Tock's design, our experiences using Rust (both correctly and incorrectly), and how the project has benefited from and served both academia and industry.

We hope the technical details and experiences will guide operating systems builders who are considering using Rust, and inform a pathway for continued development of a research operating system with practitioners deploying it.

## 8 Acknowledgements

We thank our shepherd and anonymous reviewers who provided valuable feedback on how to communicate the experiences and lessons in this paper.

At least 400 people have contributed to the Tock open-source project. In particular, Johnathan Van Why, Hudson Ayers, and Alexandru Radovici are additional core maintainers who participated in the project evolution described in this paper. The earliest of Tock's design iterations were encouraged and influenced by Michael Andersen, who also designed and built the hardware platform that Tock initially targetted. Niko Matsakis initially suggested the use of interior mutability for Tock's concurrency model. Dominic Rizzo, Alyssa Haroldson, Bryan Cantrill, and Bobby Reynolds encouraged commercial adoption of Tock and provided much of the feedback that informed the project's evolution.

This work was funded in part by NSF grants 2303639, 2443589, 2144940 and 1505728, 1505684 as well as a Google Research Award.